\documentclass[pra,twocolumn,titlepage,nofootinbib,amsmath,showpacs]{revtex4}%
\usepackage{amsmath}
\usepackage{amsfonts}
\usepackage{amssymb}
\usepackage{graphicx}%
\begin{document}
\title{Constraint on axion-like particles from atomic physics}
\author{S.~G.~Karshenboim}
\email{savely.karshenboim@mpq.mpg.de} \affiliation{Pulkovo
Observatory, St.Petersburg, 196140, Russia}
\affiliation{Max-Planck-Institut f\"ur Quantenoptik, Garching,
85748, Germany}
\author{V.V. Flambaum}
\affiliation{School of Physics, University of New South Wales, Sydney  2052, Australia}

\begin{abstract}
A possibility to constrain axion-like particles from precision
atomic physics is considered.
\end{abstract}
\maketitle

A possible solution of the strong CP problem led to introduction of
a light pseudoscalar particle called axion
\cite{Sweinberg,PQ,Weinberg,Wilczeck,Moody}. Stable or nearly stable
pseudoscalar particles may also be  a part of the dark matter. This
has motivated an extensive search for such particles coming from
cosmic space and Sun, also a laboratory search  for a new
spin-dependent long-range interaction which may be mediated by these
particles. The limits on the interaction constants and masses of
such particles as well as numerous references are presented, e.g.,
in the Particle Data Group issues \cite{PDT}.

The aim of this note is to study a possibility to constrain
axion-like particles from precision atomic physics in a way similar
to constraints on pseudovector particles in \cite{prl, prd,pra}.

Being extracted from the measurements of muonium, hydrogen and
deuterium hyperfine structure, that could produce limits on the
axion-like particles complementary to macroscopic and astrophysical
searches as well as to particle-physics experiments.

To avoid any model dependence, in the results presented below we do
not assume any relations between the interaction strength and  mass
of the intermediate pseudoscalar particles which have been used to
obtain the limits on the axion models presented in \cite{PDT}.  We
also do not assume that the particles are stable or long-lived (this
is assumed in the astrophysical constrains). All what we need is
that their width is smaller than their mass.


Recently light pseudovector particles were constrained by studies of
hyperfine structure of light atoms. A comparison of precision
experimental data and an advanced theory allows to produce efficient
constraint on a pseudovector particle with masse below 1 keV/$c^2$
\cite{prl,prd,pra}.

Some of those constraints are summarized in Table~\ref{T:const1s}.
The constraint comes from a study of a small exotic Yukawa-type
contribution of interaction of the electron and nuclear spins
\begin{equation}\label{ssl}
\frac{\alpha^{\prime\prime} \bigl({\bf s}_e\cdot{\bf
s}_{X}\bigr){\rm e}^{-\lambda r}}{r}
\end{equation}
where relativistic units in which $\hbar=c=1$ are applied.

\begin{table}[phtb]
\begin{tabular}{ccr}
\hline
~~~Atom~~~ &~~~~~$X$~~~~~ &  $|\alpha_0^{{\prime\prime}}|$~~~~~~  \\
 \hline
Mu & $\overline{\mu}$ & $7.6\times 10^{-16}$\\
H & $p$  &$1.6\times10^{-15}$\\
D & $d$  & $8\times10^{-15}$\\
\hline
\end{tabular}
\caption{The constraint from the $1s$ HFS intervals on a coupling
constant $\alpha^{\prime\prime}$ for a pseudovector boson
\cite{prl,prd}. The results presented are for limit on the absolute
values of the constant at the one-sigma confidence level for the
limit of small masses ($\lambda\ll \alpha m_e \simeq 3.7\;$keV),
which is related to the Yukawa radius substantially above the Bohr
radius $a_0$.\label{T:const1s}}
\end{table}

As it was pointed out by P. Fayet \cite{fayet}, in reality an
exchange by a pseudovector particle will include not only an
Yukawa-type potential (\ref{ssl}), but also a contact term. The
latter was addressed in \cite{pra}.

Similar contributions can also come from an axion exchange. They are
studied in this paper.

The constraint on axion-meditated spin-spin coupling is
\begin{equation}
\tilde\alpha_{AeX}=\frac{\alpha^{\prime\prime}_{0}(eX)}{{\cal
F}_A(\lambda/\alpha m_e)} \;,
\end{equation}
where the limits on $\alpha^{\prime\prime}_{0}(eX)$ are  presented in
Table~\ref{T:const1s}.

The profile function ${\cal F}_A$ is
\begin{equation}
{\cal F}_A(x)=-\frac{1}{3}{\cal F}_1(x)+2{\cal F}_\delta(x)\:,
\end{equation}
where \cite{prd}
\begin{equation}
{\cal F}_1(x)=\left(\frac{2}{2+x}\right)^2
\end{equation}
and
\begin{equation}
{\cal F}_\delta (x)=\frac{4}{3x^2}\;.
\end{equation}
The first term comes from the Yukawa-type interaction and the second
is from the contact term (cf. \cite{pra}).

We note, that for $\lambda \ll  \alpha m_e$, which means $x\ll1$
\begin{eqnarray}\label{constraint}
{\cal F}_1(x)&\simeq&1\nonumber\\
{\cal F}_\delta(\lambda/\alpha m_e)&\gg& 1\;,
\end{eqnarray}
and finally we find for small $\lambda$
\begin{equation}
\tilde\alpha_{AeX}=\frac{3}{8}\alpha^{\prime\prime}_{0}(eX)\frac{\lambda^2}{(\alpha
m_e)^2}\;,
\end{equation}

The most strong constraint can be set on $e\mu$ coupling via an
exchange by an axion-like particle. Combining the experimental
result \cite{mu1shfs} on the $1s$ hyperfine interval in  muonium
with theory \cite{my_rep} (see also \cite{codata2006,prd} for
details and references) we find (cf. \cite{prl,prd})
\begin{equation}
|\tilde\alpha_{Ae\mu}| \leq \bigl(2.8\times
10^{-16}\bigr)\frac{\lambda^2}{(3.7\;{\rm keV})^2}\;.
\end{equation}

Similar but weaker constraints can be set on the $ep$ and $en$
coupling (cf. \cite{prd,pra}). For the latter we take into account
that in deuterium
\[
{\bf s}_n \simeq \frac12 {\bf s}_d\;.
\]
We also note that the deuterium constraint is substantially weaker
than that for hydrogen, which allows to neglect the proton
contribution and thus we obtain for the effective $ed$ coupling
\begin{equation}
\tilde\alpha_{Aen}=2\tilde\alpha_{Aed}\;.
\end{equation}

It is useful to express the effective coupling constant due to
exchange by an axion-like particle in more fundamental terms, i.e.
in terms of a vertex of emission of such a particle by d a fermion,
which may be parameterize as
\begin{equation}\label{vertex}
x_i \frac{m_i}{v} A \langle \Psi_i | \gamma_5 | \Psi_i \rangle\;,
\end{equation}
where $i=e,\mu,p,n$, while $A$ stands for the axion-like field.
Here, $v$ is the vacuum average of the Higgs field from the standard
electroweak theory
\[
v=246\;{\rm GeV}
\]
and $x_i$ are the parameters to constrain.

Combining the parametrization (\ref{vertex}) with Eq.
(\ref{constraint}) we find
\begin{eqnarray}\label{xx}
|x_ex_X| & =&4\pi\frac{v^2}{\lambda^2}|\tilde{\alpha}_{AeX}|\nonumber\\
&=&\frac{3\pi}{2}\alpha^{\prime\prime}_{0}(eX)\frac{v^2}{(\alpha
m_e)^2}\nonumber\\
&=& 2.0\times 10^{16}\alpha^{\prime\prime}_{0}(eX)\;,
\end{eqnarray}
where the limits on $\alpha^{\prime\prime}_{0}(eX)$ are presented in
Table~\ref{T:const1s}.
For the most strong constraint that reads
\[
|x_ex_\mu|\leq 15.6\;.
\]

For the accepted axion model the existing limitations are several
orders of magnitude stronger than our constraints (see, e.g.
\cite{PDT}). However, the results strongly depend on a model
applied. Our constraint, being model independent, is complementary
to limitations obtained by other methods.

\section*{Acknowledgments}

This work was supported in part by DFG (grant GZ: HA 1457/7-1) and
RFBR ((under grant \# 11-02-91343)), Australian research Council.
The authors are grateful to Yury Gnedin, Pierre Fayet and Maxim
Pospelov for useful and stimulating discussions. The warm
hospitality of ECT*, where the work was stared, is gratefully
acknowledged. SGK is also grateful to UNSW for their hospitality and
Gordon Godfrey foundation for their fellowship.


\begin{thebibliography}{00.}

\frenchspacing
\bibitem{Sweinberg} S. Weinberg, Phys. Rev. Lett. {\bf 37}, 657 (1976).

\bibitem{PQ} R.D. Peccei and H.R. Quinn, Phys. Rev. Lett. {\bf 38}, 1440 (1977).

\bibitem{Weinberg} S. Weinberg, Phys. Rev. Lett. {\bf 40}, 223 (1978).

\bibitem{Wilczeck}  F. Wilczek, Phys. Rev. Lett. {\bf 40}, 279 (1978).

\bibitem{Moody} J. E. Moody and F. Wilczek, Phys. Rev. D {\bf 30}, 130 (1984).

\bibitem{PDT}  K. Nakamura et al (Particle Data Group) JPG  {\bf 37}, 075021 (2010).


\bibitem{prl} S. G. Karshenboim, Phys. Rev. Lett. {\bf 104}, 220406 (2010).

\bibitem{prd} S. G. Karshenboim, Phys. Rev. D {\bf 82}, 113013 (2010).

\bibitem{pra} S. G. Karshenboim, Phys. Rev. A {\bf 83}, 062119 (2011).

\bibitem{fayet} P. Fayet, private communication.


\bibitem{mu1shfs} W. Liu, M. G. Boshier, S. Dhawan, O. van Dyck, P. Egan, X. Fei, M.
G. Perdekamp, V. W. Hughes, M. Janousch, K. Jungmann, D. Kawall, F.
G. Mariam, C. Pillai, R. Prigl, G. zu Putlitz, I. Reinhard, W.
Schwarz, P. A. Thompson, and K. A. Woodle, Phys. Rev. Lett. {\bf
82}, 711 (1999).

\bibitem{my_rep}
S.\,G.~Karshenboim, Phys.\ Rep.\  {\bf 422}, 1 (2005).

\bibitem{codata2006}
P. J. Mohr, B. N. Taylor, and D. B. Newell, Rev. Mod. Phys. {\bf
80}, 633 (2008).

\end{thebibliography}
\end{document}